\documentclass[12pt]{article}
\usepackage{amsmath,amssymb,epsfig,amsfonts}
\usepackage{graphicx,subfigure}
\usepackage[usenames, dvipsnames]{color}
\usepackage[backref]{hyperref}
\usepackage{cite}
\usepackage[all]{xy}
\usepackage{verbatim} 

\makeatletter

\newcommand{\ssn}[1]{{\color{blue}  #1}}
\newcommand{\mk}[1]{{\color{red}  #1}}

\newcommand{\be}{\begin{equation}}
\newcommand{\ee}{\end{equation}}
\newcommand{\ba}{\begin{aligned}}
\newcommand{\ea}{\end{aligned}}

\def\unit{{1\kern-.65ex {\rm l}}}
\def\1{{1\kern-.65ex {\rm l}}}

\def\CC{{\cal C}}
\def\CD{{\cal D}}

\def\CS{{\cal S}}

\newcount\hour \newcount\minute
\hour=\time \divide \hour by 60
\minute=\time
\def\now{%
\ifnum \hour<13
  \ifnum \hour=0 \advance \hour by 12 \number\hour:\else \number\hour:\fi%
     \ifnum \minute<10 0\fi%
     \number\minute%
\ A.M.%
\else \advance \hour by -12 \number\hour:%
  \ifnum \minute<10 0\fi%
  \number\minute%
  \ P.M.%
\fi%
}

\makeatother

\begin{document}

\baselineskip=18pt  
\numberwithin{equation}{section}  
\allowdisplaybreaks  



%
%


\thispagestyle{empty}

\vspace*{-2cm} 
\begin{flushright}
{\tt KCL-MTH-12-05}\\
\end{flushright}

\vspace*{0.8cm} 
\begin{center}
 {\LARGE  $G$-flux and Spectral Divisors}

 \vspace*{1.4cm}

Moritz K\"untzler and Sakura Sch\"afer-Nameki\\
\vspace*{.8cm} 
{\it Department of Mathematics, King's College, London\\
The Strand, WC2R 2LS, London, England\\}
{\tt moritz.kuentzler kcl.ac.uk, ss299 theory.caltech.edu}\\
\end{center}
\vspace*{.5cm}

\noindent
We propose a construction of $G$-flux in singular elliptic Calabi-Yau fourfold compactifications of F-theory, which in the local limit allow a spectral cover description. The main tool of construction is the so-called spectral divisor in the resolved Calabi-Yau geometry, which in the local limit reduces to the Higgs bundle spectral cover. 
We exemplify the workings of this in the case of an $E_6$  singularity by constructing the resolved geometry, the spectral divisor and in the local limit, the spectral cover. The $G$-flux constructed with the spectral divisor is shown to be equivalent to the direct construction from suitably quantized linear combinations of holomorphic surfaces in the resolved geometry, and in the local limit reduces to the spectral cover flux. 

\newpage
\setcounter{page}{1} 



\tableofcontents
\newpage


\section{Introduction}

In this paper we study singular elliptically fibered Calabi-Yau fourfold compactifications of F-theory and the 
construction of $G$-flux in these geometries. Formally, the $G$-flux is defined as a $(2,2)$ form which integrates non-trivially over holomorphic surfaces and satisfies the quantization condition \cite{Witten:1996md}
\be\label{GQuantize}
 G+\frac{1}{2}c_2(\tilde Y_4) \in H^4(\tilde Y_4, \mathbb{Z}) \,,
\ee
where $\tilde{Y}_4$ is a resolution of the  Calabi-Yau fourfold. It can of course be 
constructed by brute force in terms of holomorphic surfaces in the resolved Calabi-Yau fourfold. 
As $G$-flux depends crucially on the singularity structure of the elliptic fibration, it is natural to anticipate a framework that makes more direct use of the singularity structure in the construction of the flux. 
Recent work has lead to much progress in the development of such a  framework. 

Progress has been made using various approaches: in local models flux was constructed in the usual heterotic/F-theory inspired setup of spectral covers \cite{FMW} starting with \cite{Donagi:2009ra, Marsano:2009ym, Marsano:2009gv}. 
On the other hand the resolution of a general $A_4$ singularity was proposed in \cite{EY} and used to directly construct $G$-flux in terms of homolorphic surfaces in \cite{MS}.  Other approaches include studying the Sen limit to IIB orientifolds \cite{Krause:2012yh}, construction in terms of algebraic cycles \cite{Braun:2011zm} and M/F-theory duality \cite{1109.3191, Intriligator:2012ue}\footnote{For an overview of relatively recent developments in the field of F-theory compactifications on elliptic Calabi-Yau fourfolds see \cite{Weigand:2010wm}.}.

An approach to $G$-fluxes which makes use of the singularity structure was proposed in the papers \cite{Marsano:2010ix, Marsano:2011nn} and shown to be consistent with the direct construction of the flux in  $\tilde{Y}_4$ in \cite{MS}. The idea is to construct the fluxes from a special divisor, the  {\it spectral} (or Tate) {\it divisor}, in the 
resolved Tate form \cite{Bershadsky:1996nh, Katz:2011qp} of the geometry, $\tilde{Y}_4$, which behaves close to the singularity in the same way as the spectral cover of the Higgs bundle in the local model. 
This proposal was exclusively performed and tested in the context of $A_4$ singularities.  

In this paper we point out that this spectral divisor formalism generalizes to all singularity types, which allow for a local spectral cover description as explained in \cite{FMW}. However, to make contact with the local Higgs bundle spectral cover, the Tate form has to be modified, as we  explain in the next section. We then exemplify this construction in the case of $SU(3)$ covers, which correspond to a singularity of type $E_6$.  In section \ref{sec:generalities} we construct the resolution of the $E_6$ singularity, and for completeness determine the higher-codimension structure of the singularity. In section \ref{sec:G-flux}, the resolution is used to construct the properly quantized $G$-flux, which preserves the $E_6$ symmetry, both directly and using the spectral divisor formalism. Both approaches agree and in the local limit give rise to a consistent local spectral cover flux.

\newpage

\section{$G$-flux and Spectral divisor}
\label{sec:SpecDiv}

\subsection{Spectral Form of the Singularity}

Consider a singular elliptic Calabi-Yau fourfold $Y_4$ with base three-fold $B$ and with a singularity of type $G$ along a surface $S$, given by $z=0$ in terms of a local holomorphic coordinate $z$ on $B$. 
The equation for $Y_4$ can then be put globally into the Tate form for $G$ \cite{Bershadsky:1996nh} (modulo subtleties discussed in \cite{Katz:2011qp})
\be\label{TateForm}
y^2 + a_1 xy + a_3 y = x^3 + a_2 x^2 + a_4 x + a_6  \,,
\ee
where the vanishing order in $z$ is determined by the type of the singularity
\be
a_i = z^{n_i} b_i \,,
\ee
where $b_i$ are sections of $\mathcal{O}(i c_1 - n_i S)$ and $c_1= c_1 (B)$.
Consider F-theory on $Y_4 \times \mathbb{R}^{1,3}$, then the physics close to the locus $z=0$ has a description in terms of an ${\mathcal{N}}=1$ supersymmetric gauge theory with gauge group $G$.

We will restrict out attention to gauge groups $G$, which can be 
thought to arise from higgsing of an underlying $E_8$ gauge theory by adjoint scalar vevs, and where the data of the gauge theory is geometrically encoded in a spectral cover $\mathcal{C}$ over $S$ \cite{FMW}. 
Additional data corresponding to $G$-flux is  encoded in spectral cover fluxes, which are constructed from line bundles over $\mathcal{C}$.  
This construction has a dual description, in case the CY fourfold has a $K3$ fibered structure, to heterotic compactifications with $H=SU(N)$ or $Sp(N)$ vector bundles, where $H$ is the commutant of $G$ inside $E_8$. 
We will restrict our discussion to the case when such a spectral cover (SC) construction is known to exist in the local limit, and denote these groups by type $G_{\rm SC}$. Concretely, 
the cases that allow for a SC formulation in the local limit have vanishing orders $n_i$ of the sections $a_i$ and the discriminant $\Delta$ for the elliptic fibration that are summarized in the following table
\be\label{Gs}
\begin{array}{c|c|c|c|c|c|c|c}
G_{\rm SC}  &H & n_1 & n_2 & n_3 & n_4 & n_6 & \Delta \cr\hline
E_7 &SU(2)& 1 & 2 & 3 & 3 & 5 & 8 \cr
E_6 &SU(3)& 1& 2 & 2 & 3 & 5 & 8 \cr
SO(10)&SU(4)& 1& 1& 2& 3& 5 & 7 \cr
SU(5) &SU(5)& 0 & 1 & 2 & 3 &5 &5 \cr\hline
SO(11) &Sp(2)& 1&1&3&3&5&8  
\end{array}
\ee
There are of course are other groups that can arise by a higgsing of an $E_8$ gauge theory. However, the commutant $H$ of $G$ is then not of $SU(N)$ or $Sp(N)$ type, and so the construction of fluxes will not come from a SC (see \cite{FMW}).

In concrete F-theory constructions, in particular in view of phenomenologically relevant models, we often require $U(1)$ symmetries in addition to the gauge symmetry $G$. Realization of these in the spectral cover formalism have been shown to be possible by imposing a factored form for the spectral cover \cite{Tatar:2009jk, Marsano:2009gv, Marsano:2009wr, Grimm:2010ez, Dolan:2011iu, Krause:2011xj}. Gauge fluxes in the direction of these $U(1)$s have been constructed from the factored spectral cover. One important question is then, how these local constructions lift to the full Calabi-Yau fourfold $Y_4$, and its resolution $\tilde{Y}_4$, and how fluxes associated to $U(1)$ symmetries are realized in this context. 

In \cite{Marsano:2010ix, Marsano:2011nn, MS} a proposal was made in terms of spectral divisors, which in a local limit reduce to the spectral cover $\mathcal{C}$ of the Higgs bundle. 
The construction there was mainly focused on the settting of $G=SU(5)$. 
We will detail how this proposal for a spectral divisor formalism generalizes for any gauge group $G$, which allows for a spectral cover construction in the local limit. 

Recall that in \cite{Marsano:2010ix, Marsano:2011nn} the Tate divisor was defined as the divisor that in the local limit reduces to the spectral cover, with the property that in the presence of additional $U(1)$ symmetries it maintains the factored form of the spectral cover. In the resolved Tate form $\tilde{Y}$ for $SU(5)$ it can be characterized by the equation
\be\label{SU5Tate}
\mathcal{C}_{\text{Tate}, SU(5)}:\qquad x^3 = y^2\,.
\ee
The local limit is defined by taking 
\be\label{LocLim}
t= x/y\rightarrow 0\,,\qquad \hbox{while} \qquad s= z/t \quad \hbox{fixed} \,.
\ee
Indeed,  the Tate divisor reproduces in this local limit  in the case of $SU(5)$ the spectral cover \cite{Donagi:2009ra}, i.e.
\be
\mathcal{C}_{{\text{SC}}, SU(5)}: \qquad  b_1 - b_2 s + b_3 s^2 - b_4 s^3 - b_6 s^5 =0 \,.
\ee
More generally, the definition of the Tate divisor has to be refined\footnote{We will refer to the divisor, which in the local limit results in the spectral cover, maintaining potential factorizations, as {\it the spectral divisor}. As in general, this will not result from the Tate form, we will not use the terminology {\it Tate divisor}.}.

Applying the characterization in terms of  (\ref{SU5Tate}) and the limit (\ref{LocLim}) for a general Tate form yields
\be
b_1 s^{n_1} t^{n_1+5}-b_2 s^{n_2} t^{n_2+4}+b_3 s^{n_3} t^{n_3+3}-b_4 s^{n_4} t^{n_4+2}-b_6 s^{n_6} t^{n_6} =0\,.
\ee
On the other hand, the spectral cover equations for the groups  in (\ref{Gs}) are
\be\label{CSC}
\begin{array}{c|c|l}
G   &H & {\cal{C}}_{\text{SC}} \cr\hline
E_7 &SU(2)&    \bar{b}_6 s^2+\bar{b}_4  \cr
E_6 &SU(3)& \bar{b}_6 s^3+\bar{b}_4 s-\bar{b}_3 \cr
SO(10) &SU(4)&  \bar{b}_6 s^4+\bar{b}_4 s^2-\bar{b}_3 s+\bar{b}_2 \cr
SU(5) &SU(5)& \bar{b}_6 s^5+\bar{b}_4 s^3-\bar{b}_3 s^2+\bar{b}_2 s-\bar{b}_1  \cr\hline
SO(11) &Sp(2)& \bar{b}_6 s^4+ \bar{b}_4 s^2+\bar{b}_2  \cr
\end{array}
\ee
Here the sections $\bar{b}_n = b_n|_{S}$. 
Each of these arise from $y^2= x^3$ in the local limit (\ref{LocLim}) as the leading equations in $t$. However, in order to define the lift into the resolved geometry $\tilde{Y}$ this is not a suitable  definition of the spectral divisor. 
Consider what we will refer to as the {\it spectral form} of the singular elliptic CY, namely, each of the Tate forms can be put into the following spectral form by shifting the coordinates $x$ and $y$. This form has appeared as we realized recently in \cite{Grimm:2010ez}.  

For $E_7$, we can shift successively 
\be
y\rightarrow y  - {1\over 2}\left(b_1 z   x  +b_3 z^3\right)  \,,\qquad  x\rightarrow x-{1\over 12} z^2 (b_1^2 + 4 b_2)
\ee 
so that the equation in the new coordinates takes the form
\be\label{E7Spectral}
 y^2  = x^3 +  b'_4 z^3 x + b'_6 z^5  \,,
\ee
with new sections $b_n'$.
Note that this equation satisfies the requirements from Kodaira's classification for an $E_7$ singular fiber at $z$, i.e. the corresponding Weierstrass form $y^2 = x^3 + f x + g$ satisfies that the degrees of vanishing at $z$ are 
\be
\hbox{deg}(f) =3 \,,\qquad \hbox{deg} (g) \geq 5 \,,\qquad \hbox{deg} (\Delta) = 9 \,.
\ee
In the form (\ref{E7Spectral}), which we will refer to as the spectral form of the $E_7$ singularity, we can now define the spectral divisor $\mathcal{C}_{\text{spectral}}$ by
$y^2 = x^3$, which under (\ref{LocLim})  limits precisely to the spectral cover for the  $E_7$ gauge theory. 

For each of the cases in (\ref{Gs}) we can pass from the Tate form to a unique spectral form\footnote{This is unique in the sense that it has the minimal set of non-vanishing sections $b_i$, which give rise to the required degrees of vanishing in the Kodaira classification for singular elliptic fibers.}
\be
\ba
E_6: &\qquad y\rightarrow   y - {1\over 2} b_1 z  x  \,,\qquad  x\rightarrow x + {1\over 12}  z^2 (b_1^2 + 4 b_2) \cr
SO(10):& \qquad  y\rightarrow y - {1\over 2} b_1 z  x \cr
SO(11):& \qquad  y\rightarrow y  - {1\over 2}\left(b_1 z   x  +b_3 z^3\right)   \,.
\ea
\ee
For $SU(5)$ the Tate form is conveniently already the spectral form. The resulting spectral forms of the singularities are in summary
\be
\begin{array}{c|rl}
G   & \text{Spectral form}&\!\!\!\text{of singularity}\cr\hline
E_7 &    y^2  &= x^3 +  b_4 z^3 x + b_6 z^5   \cr
E_6 &     y^2+ b_3 z^2 y   &= x^3 +   b_4 z^3 x + b_6 z^5   \cr
SO(10) &   y^2+ b_3 z^2 y  & = x^3 + b_2 z x^2 +   b_4 z^3 x + b_6 z^5 \cr
SU(5) &     y^2+b_1 x y + b_3 z^2 y &  = x^3 + b_2 z x^2 +   b_4 z^3 x + b_6 z^5  \cr\hline
SO(11) &   y^2   & = x^3 + b_2 z x^2 +   b_4 z^3 x + b_6 z^5 \cr
\end{array}
\ee
In the spectral form of the singularity we can now define the {\it spectral divisor},  i.e. the divisor which in the local limit (\ref{LocLim}) reduces to the spectral cover of the Higgs bundle, and furthermore maintains any factored form of the spectral cover\footnote{To eliminate any confusion in terminology: this is what in the $SU(5)$ case was named Tate divisor, however, for obvious reasons this is not a suitable name since the Tate form is not relevant for this discussion. In 
\cite{Marsano:2010ix, Marsano:2011nn} spectral divisors were defined as the family of divisors in the resolved fourfold, that limit locally to the spectral cover. The member of this family, which furthermore lifts a factored form of the spectral cover is the most relevant for the purpose of constructing fluxes (in particular $U(1)$ fluxes corresponding to the factorization of the spectral cover). Since this is the key object to study, it will be refered to as  {\it the spectral divisor}.}, in terms of the equation in the spectral form by 
\be\label{SpecDef}
\mathcal{C}_{\text{spectral}}:\qquad y^2 = x^3 \,.
\ee
In the local limit defined as in (\ref{LocLim}), it is straightforward to see that the spectral divisor restricts to the SC of the local models.


\subsection{Local $G$-flux from Spectral Covers}
\label{sec:LocalGfluxfromSC}

Before discussing the construction of global flux from the spectral divisor, it is useful to recall the construction in the local model. 
In the local framework of spectral covers, flux is constructed as follows (see \cite{Donagi:2009ra} and for a summary appendix D of \cite{MS}). Consider
\be\label{LocFlux} 
  \mathcal{C}_{\text{SC}} \cdot \pi^* \Sigma \qquad \hbox{and} \qquad \mathcal{C}_{\text{SC}} \cdot \sigma_{\text{SC}} \,,
\ee
where $\sigma_{\text{SC}}$ is the class of the hyperplane of the $\mathbb{P}^1$-bundle $Z=\mathbb{P} (\mathcal{O} \oplus  K_S)$ in which the spectral cover is embedded, and $\Sigma$ is a curve in $S$ and 
$\pi$ the projection map 
\be
  \pi: \quad Z\ \rightarrow \ S\,.
\ee
The thereby induced covering map of the spectral cover will be denoted by 
\be
  p: \quad \mathcal{C}_{\text{SC}} \ \rightarrow\ S \,.
\ee
To describe the gauge bundle in a local model, we specify a line bundle $L$ on $\mathcal{C}_{\text{SC}}$ (\ref{CSC}), which via the pushforward gives rise to an $H$-gauge bundle. 
For $H=SU(N)$ we require tracelessness, which amounts to
\begin{equation}
  c_1(p_{*}L)=p_{*}c_1(L)-\frac{1}{2}p_{*}r=0 \,,
  \label{eq:LocalGaugeQuantizationCondition}
\end{equation}
where $r$ denotes the ramification divisor of the covering $p$ and is given by
\be  \label{eq:RamificiationDivisor}
  r=\left(\mathcal{C}_{\text{SC}}-\sigma_{\text{SC}}-\sigma_\infty\right) \cdot {\mathcal{C}_{\text{SC}}}
    = ((N-2) \sigma_{\text{SC}}+{\pi^*(\eta-c_1{(S)})}) \cdot \mathcal{C}_{\text{SC}} \,.
\ee
We used the standard shorthand $\sigma_\infty=\sigma_{\text{SC}}+\pi^*c_1(S)$. The class $\eta$ is defined via 
\be
[\mathcal{C}_{\text{SC}} ] = N\sigma_{\text{SC}} + \pi^* \eta \,.
\ee 
The tracelessness condition (\ref{eq:LocalGaugeQuantizationCondition}), which amounts to requiring that the projection of the spectral flux to $S$ is trivial, leaves only a specific combination of the two types of local spectral fluxes for the $SU(N)$ case (for $SU(5)$ this was obtained in \cite{Donagi:2009ra}, and for split covers in \cite{Marsano:2009gv, Marsano:2009wr})
\be\label{GammaFlux}
\gamma = \alpha (N \sigma_{\text{SC}} - \pi^* (\Sigma_N)) \cdot \mathcal{C}_{\text{SC}} \,, \qquad \alpha \in \mathbb{C} \,.
\ee 
The curve $\Sigma_N$ is characterized by $b_{N_b}=0$ in (\ref{CSC}), where $b_{N_b}=0$ corresponds to the class of the curve $s=0$ in the SC, i.e. 
\be
\Sigma_N =  (\eta - N_b  c_1 {(S)})\,,
\ee
with $(N, N_b)=(2, 4), (3, 3), (4, 2), (5, 1)$. So, the following combination has   to be an 
integral class
\be
\frac{r}{2}+\gamma= { \left(-1+ N \alpha+\frac{N}{2}  \right)\sigma_{\text{SC}}}+\pi^*\left( \left(\frac{1}{2}-\alpha\right) \eta - \left(\frac{1}{2}-\alpha N_b\right) c_1(S) \right) \,.
\ee
For odd $N$ this flux is properly quantized  by choosing $\alpha\in\mathbb{Z}+\frac{1}{2}$. However, for $N$ even, such as in the case of $SO(10)$ singularities, the universal flux is not automatically properly quantized, unless there are further assumptions about $S$ (e.g.\ $c_1(S)$ even).

\subsection{Global $G$-flux from Spectral Divisors}
\label{sec:GlobalGfluxfromSC}

The discussion in the last section defines a  divisor in the spectral form for the singularities of type $G_{\rm SC}$ which we can now use to carry out the construction of global $G$-fluxes etc, as outlined in \cite{Marsano:2010ix, Marsano:2011nn, MS}. The direct construction using holomorphic surfaces in the resolved geometry can be connected to the construction with the spectral divisor, as was demonstrated for $SU(5)$ in \cite{MS}, and as we will show for $E_6$ in the following. The flux constructed in this way is quantized \cite{Witten:1996md} by means of the second Chern class of the resolved geometry
\be
  G+\frac{1}{2}c_2(\tilde Y_4) \in H^4(\tilde Y_4, \mathbb{Z}) \,.
\ee

Let $\tilde{Y}_4$ be the resolution of the singular Calabi-Yau fourfold $Y_4$, where at least all codimension 1 singularities have been blown up, for instance along the lines of \cite{EY, MS, LS}. The resolution is usually done starting with the Tate form of the singular fourfold. However, likewise, we can pass to the spectral form, which is what we will consider\footnote{In practice this amounts to setting some of the coefficients in the Tate form to 0.}. 
The proper transform of $\mathcal{C}_{\text{spectral}}$ will generically be reducible, with components correponding to exceptional divisors of the blow-ups, and we refer to the spectral divisor in the resolved geometry as the irreducible component of this, after subtraction of various exceptional divisors.

To make contact between the local SC construction and the global $G$-flux obtained from linear combinations of surfaces, consider  the surfaces in the resolved fourfold $\tilde{Y}_4$ obtained from divisors $D$ in $B$ that restrict to curves $\Sigma$ in $S$
\be\label{SpectralFluxTypes}
\mathcal{S}_{D}  = \mathcal{C}_{\text{spectral}} \cdot D \qquad  \hbox{and }\qquad 
\mathcal{S}_{\sigma_{\text{SC}}}   \,.
\ee
These are the global analogs of (\ref{LocFlux}). 
The surface $\mathcal{S}_{\sigma_{\text{SC}}}$ is defined to contain in the local limit the matter curve that is defined in the spectral cover by $\sigma_{\text{SC}} \cdot \mathcal{C}_{\text{SC}}$, which amounts to $s=0$ inside $\mathcal{C}_{\text{SC}}$ in (\ref{CSC}), i.e. the {\bf 10} matter curve $b_1=0$ for $SU(5)$, the ${\bf 27}$ matter curve $b_3=0$ for $E_6$ etc. 

The lift of the universal spectral cover flux (\ref{GammaFlux}) requires  the special case when $D$ is 
\be
\mathcal{S}_{p^*(\eta - N_b c_1{(S)})} =  \mathcal{C}_{\text{spectral}} \cdot p^*(\eta - N_b c_1{(S)})\,.
\ee
Only a linear combination of these will be the lift of a traceless local flux and does not break the symmetry with respect to the group $G$, i.e. intersects trivially with the Cartan divisors of the resolved geometry.  
The ramification divisor lifts to the surface
\be
\mathcal{S}_{r} = (N-2) \mathcal{S}_{\sigma_{\text{SC}}} + \mathcal{S}_{p^*(\eta - c_1{(S)})} \,.
\ee
The properly quantized flux, that is the global lift of the universal flux for odd $N$, is then given by
\be\ba
   G_{\text{spectral}} &= \frac{1}{2} (2n+1)  \left( N \CS_{\sigma_{\text{SC}}} - \CS_{p^*(\eta - N_b c_1{(S)})} \right)  \cr
   &=  \frac{1}{2} (2n+1)  \left( N \CS_{\sigma_{\text{SC}}} - \mathcal{C}_{\text{spectral}} \cdot {p^*(\eta - N_b c_1{(S)})} \right)  \,.
   \ea
\ee
This has been explicitly confirmed for $SU(5)$ in \cite{MS}, and in the remainder of this paper, we will show this proposal works also in the case for $E_6$, which in particular has a spectral form that differs from the standard Tate form. 


\section{Example: $E_6$ Singularity}
\label{sec:generalities}

The resolutions of the Tate forms for singularities (\ref{Gs}) in Calabi-Yau fourfolds in codimensions 1, 2 and 3 have been constructed for $SU(5)$ \cite{EY, MS}, and more generally will appear in \cite{LS}. A non-trivial  example to illustrate and test our proposal for the $G$-flux construction from spectral divisors is $G=E_6$, for which the spectral form differs from the Tate form. First we consider the resolution of the $E_6$ singularity, and then construct $G$-fluxes, both directly using surfaces in the resolved CY fourfold and by making connection to the spectral divisor construction (in particular the local limit), and show the consistency of these two approaches. 

As a beneficial corollary to this we study the higher codimension structure of the elliptic fibration with an $E_6$ singularity and show how along the codimension 2 locus of enhanced symmetry the fibers split, realizing the matter in the ${\bf 27}$ of $E_6$. Furthermore in codimension 3, the Yukawa interaction ${\bf 27}\times{\bf 27}\times{\bf 27}$ is shown to be generated, as three matter divisors in the ${\bf 27}$ become homologous. This confirms the logic put forward in \cite{MS}, that although the fibers in codimension 3 may not have intersection relations governed by the  Dynkin diagrams of higher rank gauge groups, this does not contradict the generation of Yukawa couplings. The existence of the latter depends on  the splitting of matter divisors in such a way, that they become homologous to each other. 

\subsection{Setup}

We consider the Tate form for $E_6$ as defined in (\ref{TateForm}, \ref{Gs}). 
As in \cite{EY, MS}, we construct the resolution in the auxiliary 5-fold
\begin{equation}
  X_5=\mathbb{P}\left({\cal{O}}\oplus K_{B}^{-2}\oplus K_{B}^{-3}\right) \,,
\end{equation}
i.e. $X_5$ is a $\mathbb{P}^2$ bundle over the base of the elliptic fibration, $B$. Divisors on $X_5$ consist of pullbacks of divisors on $B$ under the projection
\begin{equation}
  \pi_X:X_5\rightarrow B
\end{equation}
and a new divisor $\sigma$ inherited from the hyperplane of the $\mathbb{P}^2$ fiber\footnote{Note that $\sigma$ differs of course from the divisor $\sigma_{\text{SC}}$ which we introduced in section \ref{sec:LocalGfluxfromSC}. The same applies to $c_1$, a shorthand for $\pi_X^*(c_1(B))$ we use in the following that differs from $c_1(S)$ used in sections \ref{sec:LocalGfluxfromSC} and \ref{sec:GlobalGfluxfromSC}.}.
The projective coordinates $w$, $x$, and $y$ on the $\mathbb{P}^2$ fiber of $X_5$ have the following classes in $X_5$
\be   \label{wxysections}
\ba
   {[w]} &= \sigma \,,\qquad 
  \  \  [x] = \sigma+2c_1\,,\quad 
    \,  [y] = \sigma+3c_1\,,\cr
    [z] &= S \,,\qquad   [a_m] = mc_1 \,,\qquad [b_m] = mc_1 - \mathrm{deg}(a_m)S  \,.
\ea
\ee
Here, $z$ is the section that vanishes along $S$, which is the component of the discriminant with the singularity of type $E_6$. The general Tate form is
\begin{equation}
y^2 + a_1 x y + a_3 y  = x^3 + a_2 x^2 + a_4 x + a_6\,,
\end{equation}
which for an $E_6$ singularity at $z=0$ specializes to
\begin{equation}
  \text{deg}(a_1) =1 \,,\qquad 
  \text{deg}(a_2) =2 \,,\qquad 
  \text{deg}(a_3) =2 \,,\qquad 
  \text{deg}(a_4) =3 \,,\qquad 
  \text{deg}(a_6) =5 \,,
\end{equation}
i.e. inside $X_5$ in homogeneous coordinates this is
\begin{equation}
  y^2 w + b_1 z x y w + b_3 z^2 y w^2  =  x^3 + b_2 z^2 x^2 w + b_4 z^3 x w^2 + b_6 z^5 w^3 \,. 
  \label{TateE_6}
\end{equation}
The discriminant has the following expansion in $z$
\begin{equation}
  \begin{aligned}
    \Delta =& -27 b_3^4 z^8\cr
    & +\left(\left(b_1 b_3+2 b_4\right) \left(\left(b_1^2+36 b_2\right) b_3^2-32 b_1 b_4
    b_3-32 b_4^2\right)-216 b_3^2 b_6\right) z^9+O\left(z^{10}\right) \,.
  \end{aligned}
\end{equation}
In codimension 2, i.e. the first subleading order in $z$, the only locus of symmetry enhancement (corresponding to the matter curve in the local description) is
\be
  b_3=0 \,.
\ee
The codimension 3 locus of enhanced symmetry, i.e. the Yukawa interaction, arises at 
\be
  b_3=b_4=0 \,.
\ee
 
\subsection{Resolution of the $E_6$ Singularity}

We will resolve the singularity in the Tate form. As will be made clear in the discussion of $G$-fluxes, the resolution can be easily obtained from this for the spectral form of section \ref{sec:SpecDiv}. The resolution in the spectral form of section \ref{sec:SpecDiv} proceeds in exactly the same way and can be recovered from the following by setting $b_1=b_2=0$. Most importantly, all homological relations between the various divisors, which are crucial for the construction of $G$-fluxes, carry over unaltered.

\subsubsection{Resolution in codimension 1}

First resolve the geometry in codimension 1. The geometry is singular along
\be
x=y=z=0 \,,
\ee
along which we blow up 
by introducing a $\mathbb{P}^2$ with projective coordinates $[x_1, y_1, \zeta_1]$, which are related to the original coordinates by
\begin{equation}
\hbox{Blow-up 1:} \quad  x = \zeta_{1} x_1 \,, \qquad y = \zeta_{1} y_1 \,,\qquad z = \zeta_{1} z_1\,,
\end{equation}
where $\zeta_{1}=0$ gives rise to an exceptional divisor $E_1$. We repeat this process along all the codimension 1 singular loci
\be
\ba  
\hbox{Blow-up 2:} \quad &x_1 = x_{2}\zeta_{2} \,,\qquad y_1 = y_2\zeta_{2} \,,\qquad \zeta_{1} = \zeta_{12}\zeta_{2} \cr
\hbox{Blow-up 3:} \quad& y_2=y_3\zeta_{3} \,, \qquad\,  \zeta_2=\zeta_{123} \zeta_{3} \,, \quad\  \zeta_2=\zeta_{23} \zeta_{3} \cr
\hbox{Blow-up 4:} \quad&  y_3=y_4 \zeta_{4} \,, \quad \ \,  \zeta_{123}=\zeta_{1234}\zeta_{4} \,, \quad \zeta_{3}=\zeta_{34} \zeta_{4} \,,
\ea
\ee
where each blow-up gives rise to an exceptional divisor $E_i$ specified by $\zeta_i=0$. After proper transforming the resulting equation, the fourfold, which is now resolved in codimension 1, takes the form
\be\label{E6BU}
\ba   
 0= & -\zeta_{23}^2 \zeta_{34} x_{2}^3 \zeta_{1234} + w \left[ y_4^2 + y_4 z_1 (\zeta_{23} b_1 \zeta_{34} \zeta_{4} x_{2} + b_3 w z_1)\zeta_{1234} \right]  \cr
    & \qquad - w \left[ \zeta_{23} \zeta_{34} \zeta_{4} z_1^2 \zeta_{1234}^2 ( \zeta_{23} b_2 \zeta_{34} \zeta_{4} x_{2}^2 + b_4 w x_{2} z_1 + b_6 \zeta_{34} \zeta_{4}^2 w^2 z_1^3 \zeta_{1234} ) \right] \,. 
\ea  
\ee
This is now a smooth fibration in codimension 1\footnote{One can check explicitly that this is non-singular:  every combination of three of the seven sections $x_{2}$, $y_4$, $z_1$, $\zeta_{23}$, $\zeta_{34}$, $\zeta_{1234}$, $\zeta_{4}$ either violates one of the projectivity relations or the Tate form has a non-vanishing derivative with respect to it.}.

\subsubsection{Resolution in higher codimension}

The space (\ref{E6BU}) is still singular in higher codimension: setting $b_3=0$, the geometry still exhibits singularities at the loci $y_4= \zeta=0$, where $\zeta$ is one of the exceptional sections of the blow-ups. We will follow \cite{LS} to do the small resolutions, where each small resolution results in a new $\mathbb{P}^1$, characterized by a section $\delta_i$
\begin{equation}
  \begin{aligned}
    y_4=\delta_5 y_5\,, & \qquad  \  \zeta_{23}=\delta_5 \zeta_{235}\,, \\
    y_5=\delta_6 y_6\,, & \qquad  \,  \ \zeta_{34}=\delta_6 \zeta_{346}\,, \\
    y_6=\delta_7 y_7\,, & \quad \ \  \,   \zeta_{1234}=\delta_7 \zeta_{12347}\,.
  \end{aligned}
\end{equation}
The  three new exceptional divisors corresponding to $\delta_5$, $\delta_6$ and $\delta_7$, are denoted by $E_5$, $E_6$ and $E_7$. The fourfold, which is now fully resolved in all co-dimensions, is then
\be \label{eq:TateFinal}
\ba
 &w y_7 \left( \delta_5 \delta_6 \delta_7 y_7 +  \delta_7 z_1 \zeta_{12347}  \left( b_1 \delta_5 \delta_6 \zeta_{4} x_{2} \zeta_{235} \zeta_{346} + b_3 z_1 w \right) \right)\cr
    &\qquad \quad =- \zeta_{235} \zeta_{346} \zeta_{12347} \left( b_6 \zeta_{346} \zeta_{4}^3 \delta_6 \zeta_{12347}^2 \delta_7^2 w^3 z_1^5 + b_4 \delta_7 \zeta_{4} z_1^3 x_{2} \zeta_{12347} w^2 \right.\cr 
&\quad \qquad \qquad\qquad \qquad\qquad  \left.    + b_2 \delta_5 \delta_6 \delta_7 \zeta_{4}^2 z_1^2 x_{2}^2 \zeta_{235} \zeta_{346} \zeta_{12347} w + \delta_5 x_{2}^3 \zeta_{235} \right) \,,
\ea 
\ee
The classes of the various sections are listed in (\ref{SecClasses}), and the resolved fourfold is in the class
\be  [\tilde{Y}_4] = 3 \sigma + 6 c_1 - 2 E_1 - 2 E_2 - 2 E_3 - 2 E_4 - E_5 - E_6 - E_7 \,.
  \label{eq:ClassOfY4} 
\ee

\subsection{Cartan divisors}

The Cartan divisors comprise the components of  $z=0$ in the resolved geometry
\begin{equation}
  z=\zeta_{235} \zeta_{346}^2 \zeta_{4}^3 \delta_5 \delta_6^2 \delta_7 \zeta_{12347} z_1 =0 \,.
\end{equation}
We now identify these with negative simple roots of $E_6$ as well as the root  $-\alpha_0$ corresponding to the extended node of the affine $E_6$ Dynkin diagram\footnote{Note that $\zeta_{12347}=0$ and $\delta_7=0$ define the same divisor in $\tilde Y_4$.}. The classes are\footnote{In writing the locus of the  Cartan divisors in $\tilde{Y}_4$ we used the projectivity relations of the blow-ups  listed in (\ref{eq:BlowupRelations}) to set various sections that cannot vanish to 1. }

\begin{equation}
  \label{CartanComps}
  \begin{array}{r|r|c|c}
   \hbox{Defining Section} & \hbox{Locus in }\tilde{Y}_4 \quad & \text{Class in }\tilde{Y}_4 & \text{Label}\\ \hline
    z_1=0 & \delta _{7} y_7^2-\zeta _{235}^2 \zeta _{346} \zeta _{12347} x_2^3=0&S-E_1 & \CD_{-\alpha_0} \\
    \delta_5=0&-b_6 \delta _6 \zeta _{235} \zeta _{346}^2-b_4 \zeta _{235} \zeta _{346} x_2+b_3 y_7 =0 & E_5 & \CD_{-\alpha_1} \\
    \delta_6=0 &-b_4 \zeta _4 \zeta _{346}+b_3 y_7-\delta _5 \zeta _{346}=0 &E_6 & \CD_{-\alpha_2} \\
    \zeta_{4}=0 & b_3 \delta _7 \zeta _{12347} y_7-\zeta _{346} \zeta _{12347} x_2^3+\delta _6 \delta _7 y_7^2 =0&E_4 & \CD_{-\alpha_3} \\
    \zeta_{346}=0&b_3 \zeta _{12347}+\delta _{5} \delta _{6} =0& E_3-E_4-E_6 & \CD_{-\alpha_4} \\
    \zeta_{235}=0 & \delta_{5}+b_3 \zeta_{12347} =0&E_2-E_3-E_5 & \CD_{-\alpha_5} \\
    \zeta_{12347}=0 &\delta_{7}=0& E_7 & \CD_{-\alpha_6} \\
  \end{array}
\end{equation}

The labeling is consistent with the standard ordering of roots of $E_6$.
The intersection of the Cartan divisors reproduces indeed the extended Cartan matrix of $E_6$, with $z_1=0$ playing the role of the affine root, and the intersections are diagramatically depicted below:

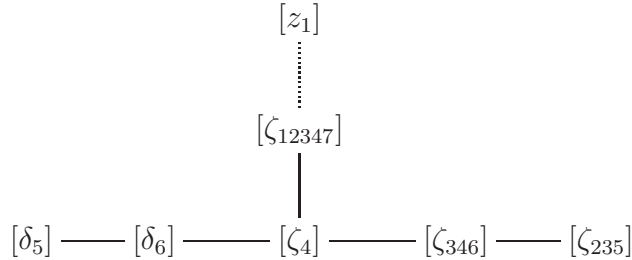
\centerline{
	\xymatrix{
		& & [z_1] & &  \\
		& & [\zeta_{12347}] \ar@{.}[u] & &  \\
		[\delta_5] & [\delta_6] \ar@{-}[l] & [\zeta_{4}] \ar@{-}[l]\ar@{-}[u] & [\zeta_{346}] \ar@{-}[l] & [\zeta_{235}] \ar@{-}[l]
	}
}

\subsection{Matter surfaces}
\label{sec:Matter divisors}

Along the codimension 2 subspace $b_3=0$ the singularity type enhances further. From the gauge theory point of view matter is generated at these loci. The intersections $\Gamma_i=[b_3]\cdot \mathcal{D}_{-\alpha_i}$ characterize the matter surfaces in $X_5$, and  we expect these  to split further such that an additional irreducible component appears in the fiber along $b_3=0$.
Indeed, as is clear from the equations for the Cartan divisors (\ref{CartanComps}) the following divisors split
\begin{itemize}
	\item $\CD_{-\alpha_4}$ splits into two components 
		\begin{equation}
			\begin{aligned}
				\Gamma_{\zeta_{346}\delta_5}: \qquad &b_3=\delta_5=\zeta_{346}=0 \cr
				\Gamma_{\zeta_{346}\delta_6}:\qquad & b_3= \delta_6=\zeta_{346} =0\,. \cr
			\end{aligned}
		\end{equation}
	\item $\CD_{-\alpha_1}$ splits into three components
		\begin{equation}
			\begin{aligned}
				\Gamma_{\zeta_{346}\delta_5}: \qquad &b_3=\delta_5=\zeta_{346}=0 \cr
				\Gamma_{-\alpha_5}:\qquad & b_3=\delta_5=\zeta_{235}=0 \cr
				\Gamma_{\delta_5b_4}:\qquad & b_3= \delta_5=b_4 x_{2}+b_6 \delta_6 =0 \,.
			\end{aligned}
		\end{equation}
	\item $\CD_{-\alpha_2}$ splits into two components
		\begin{equation}
			\begin{aligned}
				\Gamma_{\zeta_{346}\delta_6}:\qquad & b_3= \delta_6=\zeta_{346} =0 \cr
				\Gamma_{\delta_6b_4}: \qquad & b_3= \delta_6= b_4 \zeta_{4} + \delta_5=0 \,.
			\end{aligned}
		\end{equation}
\end{itemize}

With $[b_3]=3c_1-2S$, we can now determine the holomogical classes and Cartan charges of the matter divisors. 
The reducible Cartan divisors split into irreducible components that correspond to weights of the ${\bf 27}$ representation of $E_6$. Indeed, this was observed in \cite{MS} and will be explained in generality in \cite{LS}.
In detail, the charges of the irreducible matter surfaces and their identification in terms of weights of the ${\bf 27}$ as listed in Appendix \ref{sec:Weights} are
\begin{equation}\label{MatterDivisorWeights}
 \begin{array}{l|c|c}
     \hbox{Label} & \text{Cartan charges} &  E_6\text{ Weight} \\ \hline  
    \Gamma_0 & (0,0,0,0,0,1) & -\alpha_0 \\
    \Gamma_5 & (0,0,0,1,-2,0) & -\alpha_5 \\
    \Gamma_3 & (0,1,-2,1,0,1) & -\alpha_3 \\
    \Gamma_6 & (0,0,1,0,0,-2) & -\alpha_6 \\
    \Gamma_{\zeta_{346}\delta_5} & (-1,1,0,-1,1,0) & - (\mu_{\bf 27} - \alpha_1 - 2 \alpha_2 - 2 \alpha_3 - \alpha_4 - \alpha_5 - \alpha_6) \\
    \Gamma_{\zeta_{346}\delta_6} & (1,-1,1,-1,0,0) & \mu_{\bf 27} - \alpha_1 - 2 \alpha_2 - 2 \alpha_3 - 2 \alpha_4 - \alpha_5 - \alpha_6 \\
    \Gamma_{\delta_6b_4} & (0,-1,0,1,0,0) & -(\mu_{\bf 27} - \alpha_1 - \alpha_2 - 2 \alpha_3 - 2 \alpha_4 - \alpha_5 - \alpha_6) \\
    \Gamma_{\delta_5b_4} & (-1,0,0,0,1,0) &  \mu_{\bf 27} - 2 \alpha_1 - 2 \alpha_2 - 2 \alpha_3 - \alpha_4 - \alpha_6 \\
 \end{array}
\end{equation}
Adding all the weights in (\ref{MatterDivisorWeights}) together -- including multiplicities -- yields
\begin{equation}
 -\alpha_0-\alpha_1-2\alpha_2-3\alpha_3-2\alpha_4-\alpha_5-2\alpha_6,
\end{equation}
which is just the weight of the singular fiber $z=0$, as expected.

In summary we find that along $b_3=0$ the Cartan divisors corresponding to the six roots of $E_6$ split into three roots and four weights of the $\mathbf{27}$ (or $\overline{\mathbf{27}}$) representation of $E_6$. Explicitly, the divisors associated to the roots $-\alpha_0$, $-\alpha_3$, $-\alpha_5$ and  $-\alpha_6$ remain irreducible, while $-\alpha_1$, $-\alpha_2$ and $-\alpha_4$ split according to
\begin{equation}
  \begin{aligned}
    -\alpha_1 \rightarrow & -\alpha_5 + (\mu_{\bf 27} - 2 \alpha_1 - 2 \alpha_2 - 2 \alpha_3 - \alpha_4 - \alpha_6) \\
      & - (\mu_{\bf 27} - \alpha_1 - 2 \alpha_2 - 2 \alpha_3 - \alpha_4 - \alpha_5 - \alpha_6) \\
    -\alpha_2 \rightarrow & -(\mu_{\bf 27} - \alpha_1 - \alpha_2 - 2 \alpha_3 - 2 \alpha_4 - \alpha_5 - \alpha_6) \\
      & + (\mu_{\bf 27} - \alpha_1 - 2 \alpha_2 - 2 \alpha_3 - 2 \alpha_4 - \alpha_5 - \alpha_6) \\
    -\alpha_4 \rightarrow & - (\mu_{\bf 27} - \alpha_1 - 2 \alpha_2 - 2 \alpha_3 - \alpha_4 - \alpha_5 - \alpha_6) \\
      & + (\mu_{\bf 27} - \alpha_1 - 2 \alpha_2 - 2 \alpha_3 - 2 \alpha_4 - \alpha_5 - \alpha_6) \,.
  \end{aligned}
\end{equation}
We made a specific choice when resolving the higher codimension singularities, and there is in fact  a network of small resolutions, connected as in \cite{EY} by flop transitions. In particular, for each of these the fiber in codimension 2 will split into different sets of weights of the ${\bf 27}$ \cite{LS}.

\subsection{Yukawa interactions}

The codimension 3 locus of enhanced symmetry is characterized by $z=b_3=b_4=0$, along which the fourfold equation reduces to
\begin{equation}
	\begin{aligned}
		0= &\delta_5 \delta_6 \delta_7 y_7^2  -\delta_5 \zeta_{235}^2 \zeta_{346} \zeta_{12347} x_{2}^3 +b_1 \delta_5 \delta_6 \delta_7 \zeta_{47} \zeta_{235} \zeta_{346} \zeta_{12347} \zeta_{01} x_{2} y_7 \cr
	  & -b_2 \delta_5 \delta_6 \delta_7 \zeta_{47}^2 \zeta_{235}^2 \zeta_{346}^2 \zeta_{12347}^2 \zeta_{01}^2 x_{2}^2-b_6 \delta _6 \delta_7^2 \zeta_{47}^3 \zeta_{235} \zeta _{346}^2 \zeta_{12347}^3 \zeta_{01}^5 \,.
	\end{aligned}
\end{equation}
All matter surfaces remain irreducible except for 
\be	
\Gamma_{\delta_5b_4}: \qquad \delta_5 =  b_4 x_{2}+b_6 \delta_6 \zeta_{346} =0  \,,
\ee
which splits into two components in the classes 
\be
	([b_4] - [\delta_{6}]) \cdot [\delta_5] \cdot ([b_3] - [\zeta_{235} ]- [\zeta_{346}] ) \qquad \hbox{and }\qquad [\delta_6] \cdot [\delta_5] \cdot ([b_3] - [\zeta_{235} ]- [\zeta_{346}] )  \,.
\ee
Their respective Cartan charges are 
\be
	(-1, 1, 0,-1,1,0) \qquad \hbox{and } \qquad (0,-1,0,1,0,0) \,,
\ee
which are Cartan charges of other matter divisors, adding up to the Cartan charge of the corresponding matter surface $(-1, 0, 0,0,1,0) $. Thus at the locus $b_3=b_4=0$, three matter surfaces become homologous, corresponding to the Yukawa interaction
\be
\ba		(-1, 0, 0,0,1,0) \quad \rightarrow \quad &(-1, 1, 0,-1,1,0) +  (0,-1,0,1,0,0) \cr
		\mu_{\bf 27} - 2 \alpha_1 - 2 \alpha_2 - 3 \alpha_3 - \alpha_4 - \alpha_6    \quad \rightarrow \quad&
		-(\mu_{\bf 27} - \alpha_1 - 2 \alpha_2 - 2 \alpha_3- \alpha_4 - \alpha_5 -\alpha_6)  \cr
 		&-(\mu_{\bf 27} - \alpha_1 - \alpha_2 - 2 \alpha_3 - 2 \alpha_4 - \alpha_5 - \alpha_6) 
\ea
\ee	
This exactly amounts to the generation of a $\mathbf{27}\times\mathbf{27}\times\mathbf{27}$ Yukawa coupling at the $b_3=b_4=0$ locus.

\subsection{Chern classes of the resolved Fourfold}

We are interested eventually in the construction of $G$-flux satisfying the quantization condition
\begin{equation}
  G+\frac{1}{2}c_2(\tilde Y_4) \in H^4(\tilde Y_4, \mathbb{Z}) \,,
\end{equation}
for which we require the second Chern class of the resolved fourfold.
We start by working out the Chern classes of the singular fourfold $Y_4$. The total Chern class of the whole space $X_5$ is 
\begin{equation}
  c(X_5)=c(B)(1+\sigma)(1+\sigma+2c_1)(1+\sigma+3c_1)\,.
\end{equation}
The total Chern class of $Y_4$ (and especially $c_2(Y_4)$) then follows by adjunction
\begin{equation}
  \begin{aligned}
    c(Y_4) =& \frac{c(X_5)}{1+3\sigma+6c_1}\bigg|_{Y_4}=1+c_2 + 11 c_1^2  + 4 c_1 \sigma + c_3(Y_4) + c_4(Y_4) \,.
  \end{aligned}
\end{equation}
Here,  $c_i:=\pi_X^*c_i(B)$ and we used (\ref{eq:DivisorRelations}) and $\sigma\cdot_{Y_4}(\sigma+3c_1)=0$, the latter being a consequence of one of the formulae in the former.

To calculate the Chern classes of the resolved fourfold, we proceed by first calculating the Chern classes of $\tilde X_5$, using a general result from \cite{0809.2425}:
If one blows up a nonsingular subvariety $A$ which is the complete intersection of $d$ hypersurfaces $Z_1,\dots,Z_d$ of a nonsingular variety $X$ to a new subvariety $E$ obtaining a blown-up $\tilde X$, and defines the following commutative diagram

\centerline{
	\xymatrix{
		E \  \ar@{^{(}->}[r]^-{j} \ar@{->>}[d]^-{g} & \tilde{X} \ar@{->>}[d]^-{f} \\
	A\  \ar@{^{(}->}[r]^-{i} & X
	}
}
\bigskip

\noindent
then
\begin{equation}
  c(\tilde X)=\frac{(1+[E])(1+f^*[Z_1]-[E])\cdots(1+f^*[Z_d]-[E])}{(1+f^*[Z_1]) \cdots (1+f^*[Z_d])}\cdot f^*c(X) \,.
\end{equation}
As all our blow-ups and small resolutions occur along loci described by the simultaneous vanishing of several sections, we can apply this formula, with the $[Z_i]$ being the classes of these sections. The requirement that the varieties $A$ and $X$ be nonsingular does not pose a problem, since we can think of blowing up (regular) hypersurfaces in $X_5$ and passing on to $\tilde Y_4$ only after having done all the resolutions. With this, we compute the total Chern class of $\tilde X_5$ and then, with the adjunction formula, the total Chern class of $\tilde Y_4$. We obtain $c_1(\tilde Y_4)=0$, as required, and
\begin{equation}
  \begin{aligned}
    c_2(\tilde Y_4) =& c_2 + 11 c_1^2  + 13 c_1 \sigma + 3 \sigma^2 \\
    &- 4 c_1 E_1 - E_1^2 -7 c_1 E_2 + 2 E_1 E_2 - 12 c_1 E_3 + 4 E_1 E_3 + 4 E_2 E_3 + E_3^2  \\
    &- 15 c_1 E_4 + 5 E_1 E_4 + 4 E_2 E_4 + 6 E_3 E_4 + 2 E_4^2 - 6 c_1 E_5 + E_1 E_5 + 3 E_2 E_5 \\
    &+ 2 E_3 E_5 + 4 E_4 E_5 - 6 c_1 E_6 + E_1 E_6 + 3 E_2 E_6 + 3 E_3 E_6 + 3 E_4 E_6  + E_5 E_6  \\
    &- 6 c_1 E_7 + 2 E_1 E_7 + 2 E_2 E_7 + 2 E_3 E_7 + 2 E_4 E_7 + E_5 E_7 + E_6 E_7 + E_1 S \,,
  \end{aligned}
  \label{eq:c2_resolved}
\end{equation}
where all $E_i$-independent terms located in the first line correspond to $c_2(Y_4)$.

We also find the Euler character $\chi(\tilde Y_4)$ by computing the top chern class. Nicely, the result can be written as the sum of the Euler character of the singular manifold $\chi(Y_4)$ and an intersection on $S$
\begin{equation}
  \begin{aligned}
    \chi(\tilde Y_4) &= 3 \int_{B}{\left( 120 c_1^3 + 4 c_1 c_2 - 258 c_1^2 S + 183 c_1 S^2 - 42 S^3 \right)} \\
    &= \chi(Y_4) - 9 \int_{S}{\left( 86 c_1^2 - 61 c_1 S + 14 S^2 \right)} \,.
  \end{aligned}
\end{equation}
This confirms by direct computation the result conjectured in \cite{Blumenhagen:2009yv} from heterotic/F-theory duality for an $E_6$ singularity. 

\section{$G$-flux for $E_6$}
\label{sec:G-flux}

We are now in the position to construct $G$-fluxes for the $E_6$ singularity, both directly in terms of linear combination of holomorphic surfaces in $\tilde{Y}_4$, as well as using the proposal in terms of the spectral divisor and local fluxes made in section \ref{sec:SpecDiv}. 

\subsection{Direct construction  in $\tilde{Y}_4$}

\subsubsection{General conditions on $G$}

In constructing $G$-flux directly from holomorphic surfaces, we will restrict to fluxes that arise from intersections only.  
There are various conditions on the surfaces that comprise a consistent $G$-flux, in particular, 
they have to satisfy orthogonality with respect to surfaces that are pull-backs from horizontal or vertical surfaces in $Y_4$. 
Therefore, if $D$, $D_1$ and $D_2$ are pullbacks of divisors in $B$, we require
\begin{equation}
  \sigma \cdot_{\tilde Y_4} D \cdot_{\tilde Y_4} G = D_1 \cdot_{\tilde Y_4} D_2 \cdot_{\tilde Y_4} G =0 \,.
\end{equation}
This restricts us to two building blocks for $G$, namely intersections of exceptional divisors with divisors inherited from $B$ (Cartan fluxes), i.e. $E_i \cdot_{\tilde Y_4} D$ and intersections of exceptional divisors with other exceptional divisors $E_i\cdot_{\tilde Y_4} E_j$.
We furthermore want to require that the flux does not break the $E_6$ gauge symmetry, and thus has to satisfy 
\begin{equation}
  G \cdot_{\tilde Y_4} \CD_{-\alpha_i} \cdot_{\tilde Y_4} D = 0 \,.
  \label{eq:E6FluxConstraint}
\end{equation}
Both the Cartan fluxes and the pairwise intersections will intersect the Cartan divisors nontrivially and break the $E_6$ symmetry. The question is then to find linear combinations with vanishing intersections. One can check that the pairwise intersections $E_i\cdot_{\tilde Y_4} E_j$ always intersect Cartan surfaces proportional to linear combinations of
\begin{equation}
  S \cdot_{\tilde Y_4} D \cdot_{\tilde Y_4} S \qquad \mathrm{and} \qquad S \cdot_{\tilde Y_4} D \cdot_{\tilde Y_4} c_1\,.
\end{equation}
Hence, the only Cartan fluxes that can be cancelled by pairwise intersection fluxes are of the form
\begin{equation}
  E_i \cdot_{\tilde Y_4} c_1 \qquad \mathrm{or} \qquad E_i \cdot_{\tilde Y_4} S \,.
\end{equation}
This gives us an a priori 42-dimensional space. As worked out in detail in Appendix \ref{Intersection relations in X_5 and Y_4}, there are 26 divisor relations on this space. Thus, it can be parametrized with a 16-dimensional basis of surfaces. We will use the 16 surfaces
\begin{equation}
\{  E_i \cdot c_1 \,, \ E_i \cdot S \,, \  E_3 \cdot E_5 \,, \  E_3 \cdot E_6 \}\,.
\end{equation}
%

\subsubsection{Quantization of $G$}

Before evaluating the constraint (\ref{eq:E6FluxConstraint}), we check quantization of the $G$-flux (\ref{GQuantize})
with the second Chern class of the resolved manifold  (\ref{eq:c2_resolved}). The class $c_2(\tilde Y_4)$ can be rewritten by means of (\ref{eq:NonDiagonalSurfaceRelations1}, \ref{eq:NonDiagonalSurfaceRelations2}, \ref{eq:DiagonalRelations}) so that it only contains $c_2(Y_4)$ and the 16 basis surfaces
\begin{equation}
  \begin{aligned}
    c_2(\tilde Y_4) =& c_2(Y_4) + S \cdot (3E_1-E_2-E_3-E_4 + 3 (E_5 + E_6) - 4 E_7) \\
    &- c_1 \cdot (10E_1+E_2-E_4+6 E_5 + 4 E_6- 6 E_7) + E_2 \cdot E_5 - E_3 \cdot E_6 \,.
  \end{aligned}
\end{equation}
Since $c_2(Y_4)$ is an even class, we deduce that
\begin{equation}
  c_2(\tilde Y_4)= S \cdot (E_1+E_2+E_3+E_4+E_5+E_6) + c_1 \cdot (E_2+E_4) + E_2 \cdot E_5 + E_3 \cdot E_6 + \mathrm{even} \,.
\end{equation}
So a quantized general $G$-flux can be described by integers $a_i$, $b_i$, $p$, $q$ with
\begin{equation}
  \begin{aligned}
    G =& \frac{1}{2} \left(S \cdot (E_1+E_2+E_3+E_4+E_5) + c_1 \cdot (E_2+E_4) + E_2 \cdot E_5 + E_5 \cdot E_6 \right) \\
    &+ \sum_{i=1}^{7} E_i \cdot (a_i c_1 + b_i S) + p E_3 \cdot E_5 + q E_3 \cdot E_6 \,.
  \end{aligned}
\end{equation}

\subsubsection{$E_6$-invariance of $G$ and chirality}

With this form of the flux, the condition of unbroken $E_6$ gauge symmetry (\ref{eq:E6FluxConstraint}) 
can now be evaluated and the resulting solution space has three integral parameters $(a_1,b_1,N)$
\renewcommand{\arraystretch}{1.5}
\begin{equation}
  \begin{array}{ll}
    a_2=3N-a_1+1 & b_2=-2(1+N)-b_1 \\
    a_3=-3-6N-a_1 & b_3=1+4N-b_1 \\
    a_4=1+3N-a_1 & b_4=-2(1+N)-b_1 \\
    a_5=-3-6N & b_5=3+7N \\
    a_6=0 & b_6=-1-N \\
    a_7=-2a_1 & b_7=-1-2b_1 \\
    p=-2-3N & q=1+3N \,.
  \end{array}  
\end{equation}
\renewcommand{\arraystretch}{1}
This results in the $E_6$-invariant $G$-flux
\be  \label{eq:GlobalGFluxNotYetDone}
\ba
    G = &\frac{1}{2}  \left[ 3(1+2N)(c_1\cdot(E_2-2E_3+E_4-2E_5)-E_2 \cdot E_5 + E_3 \cdot E_6)\right. \cr
      &\quad - S \cdot (-E_1+3E_2-3E_3+3E_4-7E_5+E_6+2E_7 \cr
      &\qquad   \qquad\     \left. +2(2E_2-4E_3+2E_4-7E_5+E_6)N) \right]  \cr
      &+ (E_1-E_2-E_3-E_4-2E_7)\cdot(a_1 c_1 + b_1 S) \,.
\ea
\ee
Note that since $\zeta_{12347}=0$ and $\delta_7=0$ describe the same locus in the resolved geometry, the class of the last term in (\ref{eq:GlobalGFluxNotYetDone}) which is $[\zeta_{12347}] -[\delta_7]$, is equivalent to zero. Thus, $a_1$ and $b_1$ do not have any physical relevance and will cancel out of all further computations. Finally, subtracting the (homologically zero) term
$  \frac{1}{2} S \cdot (E_1-E_2-E_3-E_4-2E_7) $ from (\ref{eq:GlobalGFluxNotYetDone}), the final expression for the $G$-flux is 
\begin{equation}
  \begin{aligned}
  G = \frac{1}{2} \left(1+2N\right) [&3c_1 \cdot (E_2-2E_3+E_4-2E_5)-3 E_2 \cdot E_5 + 3 E_3 \cdot E_6 \\ 
      &- S \cdot (2 E_2 - 4 E_3 + 2 E_4 - 7 E_5 + E_6) ] \,.
  \end{aligned}
  \label{eq:GlobalGFlux}
\end{equation}
As an application, we compute the  chirality induced by this $G$-flux, which is the intersection of $G$ with the \textbf{27} matter surface $\mathbf{\CS_{\bf 27}}$ from (\ref{eq:27mattersurface})
\begin{equation}
  G \cdot_{\tilde Y_4} \mathbf{\CS_{\bf 27}}=-\frac{1}{2}(1+2N)S \cdot_{B} (6c_1-5S) \cdot_{B} (3c_1-2S) \,.
\end{equation}
Not only can this be written as an intersection in $S$, it also matches the result that one finds when computing the induced chirality in local models (cf. \ref{The gauge bundle in local $E_6$ models})
\begin{equation}
  G \cdot_{\tilde Y_4} \mathbf{\CS_{\bf 27}}=-\frac{1}{2}(1+2n) \eta \cdot_{S} (\eta-3c_1(S)) \,.
\end{equation}
The same goes for the induced D3-tadpole. For the $G$-flux in the resolved geometry, we find
\begin{equation}
  n_{\mathrm{D3,induced}}=\frac{1}{2}G\cdot_{\tilde Y_4}G=\frac{3}{8}(1+2N)^2 \, S \cdot_{B} (6c_1-5S) \cdot_{B} (3c_1-2S) \,,
\end{equation}
where the local computation (see section \ref{The gauge bundle in local $E_6$ models}) yields
\be
  n_{\mathrm{D3,induced}} = \frac{3}{8} (1+2n)^2\,  \eta \cdot_{S} (\eta-3c_1(S)) \,.
\ee
Again, the two results match.

\subsection{Local Limit and Spectral Divisor}

In this section, we relate our global description of the fourfold with local spectral cover models, and demonstrate how to use the spectral divisor formulation explained in section \ref{sec:SpecDiv}.

\subsubsection{The Spectral Divisor in the resolved Fourfold}

The spectral divisor (\ref{SpecDef}) in the resolved fourfold naively reads
\begin{equation}
  \begin{aligned}
      w^2 z_1^2 \delta_7 \zeta_{12347} \left( -b_3 y_7 + \zeta_{4} z_1 \zeta_{235} \zeta_{346} \zeta_{12347} \left( b_4 x_{2} + b_6 w \zeta_{4}^2 z_1^2 \zeta_{346} \delta_6 \delta_7 \zeta_{12347} \right) \right) & \,.
  \end{aligned}
\end{equation}
As we explained earlier, the actual spectral divisor is the irreducible component of this. The above divisor has a component $(\delta_5=0)|_{\tilde Y_4}$, as one can see from (\ref{eq:TateFinal}), and subtracting this results in the spectral divisor
\begin{equation}
  \CC_{\mathrm{spectral}}: \qquad [-b_3 y_7 + \zeta_{4} z_1 \zeta_{235} \zeta_{346} \zeta_{12347} \left( b_4 x_{2} + b_6 w \zeta_{4}^2 z_1^2 \zeta_{346} \delta_6 \delta_7 \zeta_{12347} \right)]|_{\tilde Y_4} - [\delta_5]|_{\tilde Y_4}
  \label{eq:ResolvedTateDivisor}
\end{equation}
which is in the class 
\begin{equation}
[ \CC_{\mathrm{spectral}}]= \sigma + 6c_1 - E_1 - E_2 - E_3 - E_4 - 2 E_5 - E_6 - E_7 - 2 S \,.
  \label{TateDivisorClass}
\end{equation}
For an $N$-fold spectral cover model, 
the spectral divisor should intersect with the Cartan divisors in $N$ times the weight corresponding to the representation, that in the local limit corresponds to the highest weight of a single sheet. In the case of $E_6$ this is three times the highest weight of the  ${\bf 27}$. Indeed, intersecting the spectral divisor with the Cartan divisors yields
\be
  (3,0,0,0,0,0) = 3\mu_{\bf 27} \,.
\ee

\subsubsection{Local limit and $\mathcal{C}_{\text{SC}}$}
\label{Emergence of CC_Higgs,loc}

From the singular form of the spectral divisor it is clear (by construction) that the Higgs bundle spectral cover emerges 
from the  divisor (\ref{eq:ResolvedTateDivisor}). In the resolved geometry this is less clear. To demonstrate this we first need to establish what the local limit corresponds to in $\tilde{Y}_4$ and then apply this to the spectral divisor (\ref{eq:ResolvedTateDivisor}). 

To identify the local limit, recall first that 
\begin{equation}
  \begin{aligned}
    x &= \zeta_{235}^2 \zeta_{346}^3 \zeta_{4}^4 \delta_5^2 \delta_6^3 \delta_7^2 x_{2} \\
    y &= \zeta_{235}^2 \zeta_{346}^4 \zeta_{4}^6 \delta_5^3 \delta_6^5 \delta_7^3 y_7 \\
    z &= \zeta_{235} \zeta_{346}^2 \zeta_{4}^3 \delta_5 \delta_6^2 \delta_7^2 z_1 \,,
  \end{aligned}
\end{equation}
where we replaced $\zeta_{12347}$ by $\delta_7$, as they describe the same locus in $\tilde Y_4$. The local limit parameters are then
\be
  \ba
    t&=\frac{y}{x}=\frac{y_7 \zeta_{346} \zeta_{4}^2 \delta_5 \delta_6^2 \delta_7}{x_{2}} \cr
    s&=\frac{zx}{y}=\frac{z_1 x_{2} \zeta_{235} \zeta_{346} \zeta_{4} \delta_7}{y_7} \,,
  \ea
\ee
so that the limit $t,z\rightarrow 0$ with $s=z/t$ fixed, corresponds to
\begin{equation}
  \delta_5 \delta_6 \rightarrow 0 \,.
  \label{eq:naivelocallimit}
\end{equation}
In fact (as we show in section \ref{Emergence of CC_Higgs,loc}), the proper local limit for the spectral divisor -- i.e. the one yielding the full spectral cover equation -- in the resolved geometry is $\delta_5 \rightarrow 0$. The limit $\delta_6\rightarrow 0$ on the other hand only reproduces the spectral cover equation in the patch $\delta_6=0$.

With this insight, we now apply the local limit to the spectral divisor (\ref{eq:ResolvedTateDivisor}).
In particular we will show that the restriction of the spectral divisor to $\delta_5=0$ yields the spectral cover
\begin{equation}
  \mathcal{C}_{\text{SC}}= \mathcal{C}_{\text{spectral}}\cdot_{\tilde Y_4}[\delta_5] \,.
\end{equation}
The blow-up relations (\ref{eq:BlowupRelations}), with $\delta_5$ set to zero, imply 
that the equations for the spectral divisor and the Calabi-Yau fourfold can be reduced to
\be
\ba
    0 &= \delta_5 \cr
    0 &= b_3 y_7 - \zeta_{235} \zeta_{346} \left( b_6\zeta_{346}\delta_6+b_4x_{2} \right) \,.
  \label{eq:HiggsBundleEqns2}
\ea
\ee
Note that $\zeta_{235}=0$ would imply $\delta_6 y_7=0$, which violates the blow-up relations, so that one can set $\zeta_{235}=1$. Finally, recall that the spectral divisor equation is $x^3=y^2$.
Going into the $x_{2}\not=0$ patch and plugging the spectral  divisor equation, which reduces to $y_7^2=\zeta_{346}$, into the Calabi-Yau condition, we obtain
\begin{equation}
  0=y_7 \left(-b_3+b_4 \delta_6 y_7 + b_6 \delta_6^2 y_7^3 \right) \,,
  \label{eq:WholeLocalHiggsEqn}
\end{equation}
which in the $\delta_6\neq 0$ patch, after removing a factor of $y_7$, is precisely the local equation for the SC
\begin{equation}
  \CC_{\text{SC}}: \qquad 0 = -b_3+b_4 y_7 + b_6 y_7^3 \,.
\end{equation}
For $\delta_6=0$, (\ref{eq:WholeLocalHiggsEqn}) simply gives
\begin{equation}
  0=-b_3 y_7 \,.
  \label{eq:delta6patchSC}
\end{equation}
This should describe the spectral cover in the $\delta_6=0$ patch. We now check that this is consistent with restricting the spectral divisor in the resolved geometry to $\delta_6=0$. 
Again using the blow-up relations where $\delta_6=0$ reduces the spectral divisor equation and the Calabi-Yau equation simplify to
\begin{equation}
  0 = \delta_6 = b_3 y_7 - b_4 \zeta_{346} \zeta_{4} = b_3 y_7 - b_4 \zeta_{346} \zeta_{4} - \zeta_{346} \delta_5 \,.
  \label{eq:HiggsBundleEqns1}
\end{equation}
The difference of the last two equations thus implies $\zeta_{346}=0$ or $\delta_5=0$. While the latter will be a special case of having just $\delta_5=0$, which we discussed above, the former simply yields for the spectral divisor
\begin{equation}
  0=b_3 y_7 \,.
\end{equation}
This is in fact the equation for the spectral cover we expected from (\ref{eq:delta6patchSC}).
Note, that the  \textbf{27} matter surface, which can be characterized by
\begin{equation}
  \CS_{\bf 27} = E_6 \cdot (E_3-E_4-E_6) \,,
  \label{eq:27mattersurface}
\end{equation}
meets $\CC_{\mathrm{SC}}$ exactly along the curve
\begin{equation}
  y_7=b_3=0 \,.
  \label{eq:27MatterCurve}
\end{equation}

\subsubsection{Spectral Cover flux in local $E_6$ models}
\label{The gauge bundle in local $E_6$ models}

We first construct the universal spectral cover flux for the $E_6$ model and in the next section use the spectral divisor to obtain the global version thereof. 

Spectral cover fluxes, as summarized in section \ref{sec:SpecDiv}, are constructed from line bundles over the spectral cover. 
The commutant of $E_6$ in $E_8$ is $SU(3)$, so that we are considering $SU(3)$ gauge bundles, that are obtained from push-forwards of line bundles $L$ on the spectral cover. 
Starting with the tracelessness condition (\ref{eq:LocalGaugeQuantizationCondition}), we consider a divisor $\gamma$ satisfying $p_{*}(\gamma)=0$ and write
\begin{equation}
  c_1(L)=\frac{1}{2}r+\gamma \,.
\end{equation}
Generically, $\gamma$ is a one-parameter family of divisors, given by (\ref{GammaFlux}) (for $E_6$, see also \cite{Donagi:2009ra, Chen:2010tg})
\begin{equation}
  \gamma=\alpha \left( 3 \sigma_{\text{SC}} - \pi^* (\eta-3c_1(S)) \right) \,.
\end{equation}
This flux needs to then be properly quantized. Indeed, to have  $\gamma+\frac{1}{2}r$  integral, we require $\alpha=\frac{1}{2}(2n+1)$ with an integer $n$ and thus
\begin{equation}
\gamma=\frac{1}{2}(2n+1)\left( 3 \sigma_{\text{SC}} - \pi^* (\eta-3c_1(S)) \right) \,.
\end{equation}
For completeness we compute some of the flux-related local data.  
The D3-brane charge induced by this flux is, as we already quoted above, 
\begin{equation}
  n_{\mathrm{D3,induced}} =-\frac{1}{2}\gamma\cdot_{\mathcal{C}_{\text{SC}}}\gamma=\frac{3}{8}(2n+1)^2\eta \cdot_{S} (\eta-3c_1(S)) \,.
\end{equation}
Furthermore, the chirality induced on the matter curve
\be
  [\Sigma_{\mathbf{27}}]=\mathcal{C}_{\text{SC}}\cdot\sigma_{\text{SC}}=(3\sigma_{\text{SC}}+\pi^*\eta)\cdot\sigma_{\text{SC}} 
\ee
is the intersection with the flux $\gamma$
\be
  n_{\mathbf{27}}-n_{\mathbf{\overline{27}}}=\gamma \cdot [\Sigma_{\mathbf{27}}]=-{\frac{1}{2}(2n+1)} \  \eta\cdot_{S}(\eta-3c_1(S))\,.
\ee

\subsubsection{Spectral divisor flux}

We are now ready to construct the spectral divisor fluxes, as outlined in section \ref{sec:SpecDiv}. First 
 construct surfaces that correspond to curves inside $\mathcal{C}_{\text{SC}}$, following the procedure outlined in \cite{Marsano:2011nn, MS}. There are two types of surfaces, given in (\ref{SpectralFluxTypes}): one arises from intersecting $\mathcal{C}_{\rm spectral}$ with $\sigma$, the other corresponds to $p^*D$, where $D$ intersects $S$ in a curve $\Sigma$ (which as explained in the last subsection, can be used to engineer spectral cover fluxes) and $p$ is the projection map
\begin{equation}
  p:\quad  \mathcal{C}_{\text{SC}}\  \rightarrow \ S\,.
\end{equation}
Subtractions have to be made from the fluxes in  (\ref{SpectralFluxTypes}) in order to make them orthogonal to all horizontal and vertical divisors. Solving this condition results in
 \begin{equation}
 \ba
  \CS_{p^*D} 
  &= \mathcal{C}_{\text{spectral}}\cdot D - (\sigma + 6c_1 - 2 S) \cdot D  \cr
  &= -(E_1 + E_2 + E_3 + E_4 + 2 E_5 + E_6 + E_7)\cdot D \,.
  \label{eq:CS_p^*D}
  \ea
\end{equation}
Next, consider $\CS_{\sigma_{\text{SC}}}$. This should be a surface that contains (\ref{eq:27MatterCurve}) inside $\delta_5=0$. Such an object is $\delta_6=\zeta_{346}=0$. This, though, has non-zero intersections with Cartan surfaces $\CD_{-\alpha_i}\cdot D$ other than $\alpha_1$, hence its Cartan charge differs from $\mu_{\bf 27}$. We are able to correct this using other Cartan fluxes though, so the surface class which we identify with $\CS_{\sigma_{\text{SC}}}$ is
\begin{equation}
\ba
  \CS_{\sigma_{\text{SC}}}
  &= [\delta_6] \cdot_{\tilde Y_4} [\zeta_{346}] + [b_3] \cdot_{\tilde Y_4} \left(E_2 + E_3 + E_7  \right)  \cr
  &= (E_3-E_4-E_6) \cdot_{\tilde Y_4} E_6 + (3c_1-2S) \cdot_{\tilde Y_4} (E_2+E_3+E_7) \,.
  \label{eq:CS_sigmacdotCC} 
\ea\end{equation}
In fact, the correction is exactly the homological class of the Cartan roots $- (\alpha_1 + 2 \alpha_2 + 2 \alpha_3 + 2 \alpha_4 + \alpha_5 + \alpha_6)$ which precisely amounts for the deviation of the Cartan charges of the matter surface $\Gamma_{\zeta_{346} \delta_{6}}$ in (\ref{MatterDivisorWeights}) from $\mu_{\bf 27}$. Using (\ref{eq:CS_p^*D}) and (\ref{eq:CS_sigmacdotCC}), the traceless $G$-flux with 
\be
D= p^* (\eta - 3c_1(S)) = p^*(3c_1-2S) 
\ee
that corresponds to the universal flux $\gamma$
obtained in section \ref{The gauge bundle in local $E_6$ models} is 
\begin{equation}
  \begin{aligned}
    G_{\text{spectral}} = \frac{1}{2} (2n+1) & \left( 3 \CS_{\sigma_{\text{SC}}} - \CS_{p^*(3c_1-2S)} \right) \\
    = \frac{1}{2} (2n+1) & ( 3 E_2 \cdot E_5 - 3 E_3 \cdot E_6 + 3c_1 \cdot (E_1 - 2 E_2 + E_3 - E_4 + 2 E_5 - 2 E_7) \\
    &+ S \cdot ( -2 E_1 + 4 E_2 - 2 E_3 + 4 E_4 - 7E_5 + E_6 - 4 E_7 )) \,,
  \end{aligned}
  \label{eq:LocalGFluxNotYetDone}
\end{equation}
where (\ref{eq:NonDiagonalSurfaceRelations1}, \ref{eq:NonDiagonalSurfaceRelations2}, \ref{eq:DiagonalRelations}) were used  in the last step. Finally, subtacting the trival class  $ [b_3] \cdot ([\zeta_{12347}]-[\delta_7])$ results in
\begin{equation}
  \begin{aligned}
    G_{\text{spectral}} = \frac{1}{2} (2n+1) & ( 3 E_2 \cdot E_5 - 3 E_3 \cdot E_6 - 3 c_1 \cdot (E_2 - 2 E_3 + E_4 - 2 E_5) \\
    &+ S \cdot ( 2E_2 - 4E_3 + 2E_4 - 7E_5 + E_6 )) \,.
  \end{aligned}
  \label{eq:LocalGFlux}
\end{equation}
This  spectral divisor flux  therefore precisely matches the result for the global $G$-flux that we constructed directly from linear combinations of surfaces in (\ref{eq:GlobalGFlux}). 

\section*{Acknowledgements}

We thank Craig Lawrie for collaboration on the $E_6$ resolution, and Joe Marsano, Natalia Saulina and Will Walters for discussions and comments on the draft. This work is partly supported by STFC. SSN thanks the Simons Center for Geometry and Physics for hospitality during the completion of this work.

\newpage
\begin{appendix}

\section{Details of the geometry of $\tilde{X}_5$ and $\tilde{Y}_4$}
\label{Intersection relations in X_5 and Y_4}

\subsection{Blow-up and Intersection relations}

The classes of the various sections after the blow-ups and small resolutions in   $\tilde{X}_5$ and $\tilde{Y}_4$ are 
\be\label{SecClasses}
  \ba
    {[x_{2}]} &=\sigma+2c_1-E_1-E_2\cr
    [y_7] &=\sigma+3c_1-E_1-E_2-E_3-E_4-E_5-E_6-E_7\\
    [z_1] &=S-E_1 \\
    [\zeta_{12347}] &=E_1-E_2-E_3-E_4-E_7 \\
    [\zeta_{235}] &=E_2-E_3-E_5 \\
    [\zeta_{346}] &=E_3-E_4-E_6 \\
    [\zeta_{4}] &=E_4 \\
    [\delta_i] &=E_i\, \qquad i= 5, 6, 7 \,.
\ea
\ee
The blow-up relations in $\tilde X_5$ are
\begin{equation}
  \begin{aligned}
    0 &= \sigma\cdot(\sigma+2c_1)\cdot(\sigma+3c_1) \\
    0 &= (\sigma+2c_1-E_1)\cdot(\sigma+3c_1-E_1)\cdot(S-E_1) \\
    0 &= (\sigma+2c_1-E_1-E_2)\cdot(\sigma+3c_1-E_1-E_2)\cdot(E_1-E_2) \\
    0 &= (\sigma+3c_1-E_1-E_2-E_3)\cdot(E_1-E_2-E_3)\cdot(E_2-E_3) \\
    0 &= (\sigma+3c_1-E_1-E_2-E_3-E_4)\cdot(E_1-E_2-E_3-E_4)\cdot(E_3-E_4) \\
    0 &= (\sigma+3c_1-E_1-E_2-E_3-E_4-E_5)\cdot(E_2-E_3-E_5) \\
    0 &= (\sigma+3c_1-E_1-E_2-E_3-E_4-E_5-E_6)\cdot(E_3-E_4- E_6) \\
    0 &= (\sigma+3c_1-E_1-E_2-E_3-E_4-E_5-E_6-E_7)\cdot(E_1-E_2-E_3-E_4-E_7) \,.
  \end{aligned}
  \label{eq:DivisorRelations}
\end{equation}


\subsection{Holomorphic surfaces}

To construct the $G$-flux directly, we need to determine an independent set of holomorphic surfaces in the resolved geometry. 
The various projectivity relations encode all the relations between the surfaces
\begin{equation}
  \begin{array}{lrl}
    \text{Original} & (x,y,z) &= (\zeta_{235}^2 \zeta_{346}^3 \zeta_{4}^4 \zeta_{12347} \delta_7 \delta_5^2 \delta_6^3 x_{2},\zeta_{235}^2 \zeta_{346}^4 \zeta_{4}^6 \zeta_{12347} \delta_5^3 \delta_6^5 \delta_7^2 y_7,\zeta_{235} \zeta_{346}^2 \zeta_{4}^3 \zeta_{12347} \delta_5 \delta_6^2 \delta_7 z_1) \\
    \text{Blow-up 1} & [x_1,y_1,z_1] &= [x_{2}\zeta_{235}\delta_5\zeta_{4}\delta_6\zeta_{346},y_7\zeta_{235}\zeta_{346}^2\zeta_{4}^3\delta_5^2\delta_6^3\delta_7,z_1] \\
    \text{Blow-up 2} & [x_{2},y_2,\zeta_2] &= [x_{2},y_7\delta_5\delta_6^2\delta_7\zeta_{4}^2\zeta_{346},\zeta_{12347} \delta_7\zeta_{346}\delta_6\zeta_{4}^2] \\
    \text{Blow-up 3} & [y_3,\zeta_{123},\zeta_{23}] &= [y_7\delta_5\delta_6\delta_7\zeta_{4},\zeta_{12347} \delta_7\zeta_{4},\zeta_{235}\delta_5] \\
    \text{Blow-up 4} & [y_4,\zeta_{1234},\zeta_{34}] &= [y_7\delta_5\delta_6\delta_7,\zeta_{12347}\delta_7,\zeta_{346}\delta_6] \\
    \text{Blow-up 5} & [y_5,\zeta_{235}] &= [y_7\delta_6\delta_7,\zeta_{235}] \\
    \text{Blow-up 6} & [y_6,\zeta_{346}] &= [y_7\delta_7,\zeta_{346}] \\
    \text{Blow-up 7} & [y_7,\zeta_{12347}]
  \end{array}
  \label{eq:BlowupRelations}
\end{equation}
In particular, the following sets of equations do not admit solutions:
\begin{equation}
 z_1=\delta_5=0\,, \qquad z_1=\delta_6=0\,, \qquad z_1=\zeta_{4}=0\,, \qquad z_1=\zeta_{346}=0\,, \qquad z_1=\zeta_{235}=0\,, \notag
\end{equation}
\begin{equation}
 x_{2}=\delta_6=0\,, \qquad x_{2}=\zeta_{4}=0\,, \qquad x_{2}=\zeta_{346}=0\,, \qquad x_{2}=\delta_7=0\,,
\end{equation}
\begin{equation}
 \delta_7=\zeta_{235}\,, \qquad \delta_7=\zeta_{346}\,, \qquad \delta_7=\delta_5\,, \qquad \delta_7=\delta_6=0\,, \notag
\end{equation}
\begin{equation}
 \zeta_{4}=\delta_5=0\,, \qquad \zeta_{4}=\zeta_{235}=0\,, \qquad \zeta_{235}=\delta_6=0\,, \qquad  \delta_5=\zeta_{12347}=0\,, \qquad \delta_6=\zeta_{12347}=0\,. \notag
\end{equation}
%
%
%
Using that $\sigma \cdot E_i =0$, this gives us a total of 19 relations (the last one being described below) on the space spanned by
\begin{equation}
  E_i \cdot E_j \, (i \neq j) \,, \qquad E_i \cdot c_1 \,, \qquad E_i \cdot S \,.
\end{equation}
As this space is 35-dimensional, it can be described by a 16-dimensional basis, which we can parametrize using the 14 intersections
\begin{equation}
  E_i \cdot c_1 \,, \qquad E_i \cdot S \,, 
  \label{eq:BasisSurfaces,part1}
\end{equation}
and two of the form $E_i \cdot E_j$ that are not linear combinations of (\ref{eq:BasisSurfaces,part1}). A convenient choice for the latter is
\begin{equation}
  E_2 \cdot E_5 \,, \qquad E_3 \cdot E_6 \,,
  \label{eq:BasisSurfaces,part2}
\end{equation}
From these, we can now derive all other intersections and obtain the following tables (we give an expression for $E_1 \cdot E_7$ below):
\begin{equation}
  \begin{array}{c||c|c|c|c}
    \cdot & E_1 & E_2 & E_3 & E_4 \\ \hline \hline
    E_1 & --- & S \cdot E_2 & S \cdot E_3 & S \cdot E_4 \\ \hline
    E_2 & S \cdot E_2 & --- & (2c_1-S) \cdot E_3 & (2c_1 - S) \cdot E_4 \\ \hline
    E_3 & S \cdot E_3 & (2c_1-S)\cdot E_3 & --- & (2c_1-S) \cdot E_4 \\ \hline
    E_4 & S \cdot E_4 & (2c_1-S)\cdot E_4 & (2c_1-S) \cdot E_4 & --- \\ \hline
    E_5 & S \cdot E_5 & E_2 \cdot E_5 & (S-E_2) \cdot E_5 & 0 \\ \hline
    E_6 & S \cdot E_6 & (2c_1-S) \cdot E_6 & E_3 \cdot E_6 & (2(S-c_1)-E_3) \cdot E_6 \\ \hline
    E_7 & E_1 \cdot E_7 & (2c_1-E_1) \cdot E_7 & (2c_1-E_1) \cdot E_7 & (2c_1-E_1)\cdot E_7
  \end{array}
  \label{eq:NonDiagonalSurfaceRelations1}
\end{equation}
\begin{equation}
  \begin{array}{c||c|c|c}
    \cdot & E_5 & E_6 & E_7 \\ \hline \hline
    E_1 & S \cdot E_5 & S \cdot E_6 & E_1 \cdot E_7 \\ \hline
    E_2 & E_2 \cdot E_5 & (2c_1-S) \cdot E_6 & (2c_1-E_1) \cdot E_7 \\ \hline
    E_3 & (S-E_2) \cdot E_5 & E_3 \cdot E_6 & (2c_1-E_1) \cdot E_7 \\ \hline
    E_4 & 0 & (2(S-c_1)-E_3) \cdot E_6 & (2c_1-E_1) \cdot E_7 \\ \hline
    E_5 & --- & (2c_1-S-E_3) \cdot E_6 & 0 \\ \hline
    E_6 & (2c_1-S-E_3) \cdot E_6 & --- & 0 \\ \hline
    E_7 & 0 & 0 & ---
  \end{array}
  \label{eq:NonDiagonalSurfaceRelations2}
\end{equation}
We can also write the diagonal entries of the tables as functions of our basis. To do this, we have to use the last three relations of (\ref{eq:DivisorRelations}) as well as four new relations. These new relations are all found along the same lines (and only valid within $\tilde Y_4$): When one puts both of the variables in the first column of the table below to zero and evaluates the Tate equation (\ref{eq:TateFinal}), one finds that the Tate equation becomes a product whose factors are such that the vanishing of any of them would violate the blow-up relation (\ref{eq:BlowupRelations}) in the second column.
\begin{equation}
  \begin{array}{c|c|c}
    \text{Non-vanishing pair} & \text{Blow-up} & \text{Relation in homology} \\
    \text{of sections in } \tilde Y_4 & \text{relation} & \\ \hline \hline
    z_1,\: y_7 & 1 & (S-E_1)\cdot(\sigma+3c_1-E_1-E_2-E_3-E_4-E_5-E_6-E_7)=0 \\
    z_1,\: x_{2} & 1 & (S-E_1)\cdot(\sigma+2c_1-E_1-E_2)=0 \\
    x_{2},\: \zeta_{12347} & 2 & (\sigma+2c_1-E_1-E_2)\cdot(E_1-E_2-E_3-E_4-E_7)=0 \\
    \zeta_{235},\: \zeta_{12347} & 3 & (E_2-E_3-E_5)\cdot(E_1-E_2-E_3-E_4-E_7)=0 \\
    \zeta_{346},\: \zeta_{12347} & 4 & (E_3-E_4-E_6)\cdot(E_1-E_2-E_3-E_4-E_7)=0
  \end{array}
  \label{eq:DiagonalRelations}
\end{equation}
Since the first of these homological relations involves $S \cdot \sigma$, which is not in our vector space, we find an alternative relation. We note that on the surface obtained by restricting $z_1=\zeta_{12347}=0$ in $\tilde Y_4$, necessarily $\delta_7=0$ follows. Vice versa, if one considers $z_1=\delta_7=0$ in $\tilde Y_4$, one automatically has $\zeta_{12347}=0$. This establishes
\begin{equation}
  (S-E_1)\cdot E_7 = (S-E_1)\cdot (E_1-E_2-E_3-E_4-E_7)
\end{equation}
or
\begin{equation}
  (S-E_1)\cdot (E_1-E_2-E_3-E_4-2E_7)=0 \,.
\end{equation}
Since it does not involve $\sigma \cdot S$, we will work with this relation.
We now have all the necessary information to rewrite the surfaces $E_i\cdot E_i$ in terms of our basis surfaces (\ref{eq:BasisSurfaces,part1}) and (\ref{eq:BasisSurfaces,part2}):
\begin{equation}
  \begin{split}
    E_1 \cdot E_1 &= S \cdot (E_1-2E_7) + 2 E_1 \cdot E_7 \,, \\
    E_2 \cdot E_2 &= 2 c_1 \cdot (E_2-E_1) + S \cdot (E_1-2E_7) + 2 E_1\cdot E_7 \,, \\
    E_3 \cdot E_3 &= 2 c_1 \cdot (E_2-E_1) + S \cdot (E_1-E_2+E_3-2E_7) + 2 E_1 \cdot E_7 \,, \\
    E_4 \cdot E_4 &= 2 c_1 \cdot (E_2-E_1+E_3-E_4) + S \cdot (E_1-E_2-E_3+2E_4-2E_7) + 2 E_1 \cdot E_7 \,, \\
    E_5 \cdot E_5 &= -3 c_1 \cdot (E_2-E_3-E_5) + S \cdot (2E_2-2E_3-3E_5) + 2 E_2 \cdot E_5 \,, \\
    E_6 \cdot E_6 &= -3 c_1 \cdot (E_3-E_4-E_6) + S \cdot (2E_3-2E_4+E_5-3E_6) - E_2 \cdot E_5 + 3 E_3 \cdot E_6 \,, \\
    E_7 \cdot E_7 &= 3 c_1 \cdot (E_1-E_2-E_3-E_4-3E_7) + S \cdot (-2E_1+2E_2+2E_3+2E_4+4E_7) + 2 E_1 \cdot E_7 \,, \\
  \end{split}
\end{equation}
Finally, we can use the difference of the first two relations of (\ref{eq:DiagonalRelations}) to write
\begin{equation}
  \begin{aligned}
    0&=(S-E_1)\cdot(c_1-E_3-E_4-E_5-E_6-E_7) \stackrel{\text{\ref{eq:NonDiagonalSurfaceRelations1}}}{=} (S-E_1)\cdot(c_1-E_7) \\
    \Rightarrow E_1 \cdot E_7 &= S \cdot E_7 + c_1 \cdot E_1 - c_1 \cdot S \,.
  \end{aligned}
\end{equation}

\newpage
\section{$E_6$ weights and roots}
\label{sec:Weights}

As a useful reference, we list the simple roots and the weights of the \textbf{27} representation of $E_6$. 
\begin{equation}
  \begin{array}{c|c}
    \text{Root vectors} & \text{Simple roots} \\ \hline \hline
    (2,-1,0,0,0,0) & \alpha_1 \\
    (-1,2,-1,0,0,0) & \alpha_2 \\
    (0,-1,2,-1,0,-1) & \alpha_3 \\
    (0,0,-1,2,-1,0) & \alpha_4 \\
    (0,0,0,-1,2,0) & \alpha_5 \\
    (0,0,-1,0,0,2) & \alpha_6 \\
  \end{array}
\end{equation}
\begin{equation}
  \begin{array}{c|c}
      \text{Weight vectors in the }\mathbf{27} & \text{Weights} \\ \hline \hline
      (1,0,0,0,0,0) & \mu_{\bf 27} \\
      (-1,1,0,0,0,0) & \mu_{\bf 27} - \alpha_1 \\
      (0,-1,1,0,0,0) & \mu_{\bf 27} - \alpha_1 - \alpha_2 \\
      (0,0,-1,1,0,1) & \mu_{\bf 27} - \alpha_1 - \alpha_2 - \alpha_3 \\
      (0,0,0,-1,1,1) & \mu_{\bf 27} - \alpha_1 - \alpha_2 - \alpha_3 - \alpha_4 \\
      (0,0,0,1,0,-1) & \mu_{\bf 27} - \alpha_1 - \alpha_2 - \alpha_3 - \alpha_6 \\
      (0,0,0,0,-1,1) & \mu_{\bf 27} - \alpha_1 - \alpha_2 - \alpha_3 - \alpha_4 - \alpha_5 \\
      (0,0,1,-1,1,-1) & \mu_{\bf 27} - \alpha_1 - \alpha_2 - \alpha_3 - \alpha_4 - \alpha_6 \\
      (0,0,1,0,-1,-1) & \mu_{\bf 27} - \alpha_1 - \alpha_2 - \alpha_3 - \alpha_4 - \alpha_5 - \alpha_6 \\
      (0,1,-1,0,1,0) & \mu_{\bf 27} - \alpha_1 - \alpha_2 - 2 \alpha_3 - \alpha_4 - \alpha_6 \\
      (0,1,-1,1,-1,0) & \mu_{\bf 27} - \alpha_1 - \alpha_2 - 2 \alpha_3 - \alpha_4 - \alpha_5 - \alpha_6 \\
      (1,-1,0,0,1,0) & \mu_{\bf 27} - \alpha_1 - 2 \alpha_2 - 2 \alpha_3 - \alpha_4 - \alpha_6 \\
      (0,1,0,-1,0,0) & \mu_{\bf 27} - \alpha_1 - \alpha_2 - 2 \alpha_3 - 2 \alpha_4 - \alpha_5 - \alpha_6 \\
      (1,-1,0,1,-1,0) & \mu_{\bf 27} - \alpha_1 - 2 \alpha_2 - 2 \alpha_3 - \alpha_4 - \alpha_5 - \alpha_6 \\
      (-1,0,0,0,1,0) & \mu_{\bf 27} - 2 \alpha_1 - 2 \alpha_2 - 2 \alpha_3 - \alpha_4 - \alpha_6 \\
      (1,-1,1,-1,0,0) & \mu_{\bf 27} - \alpha_1 - 2 \alpha_2 - 2 \alpha_3 - 2 \alpha_4 - \alpha_5 - \alpha_6 \\
      (-1,0,0,1,-1,0) & \mu_{\bf 27} - 2 \alpha_1 - 2 \alpha_2 - 2 \alpha_3 - \alpha_4 - \alpha_5 - \alpha_6 \\
      (1,0,-1,0,0,1) & \mu_{\bf 27} - \alpha_1 - 2 \alpha_2 - 3 \alpha_3 - 2 \alpha_4 - \alpha_5 - \alpha_6 \\
      (-1,0,1,-1,0,0) & \mu_{\bf 27} - 2 \alpha_1 - 2 \alpha_2 - 2 \alpha_3 - 2 \alpha_4 - \alpha_5 - \alpha_6 \\
      (1,0,0,0,0,-1) & \mu_{\bf 27} - \alpha_1 - 2 \alpha_2 - 3 \alpha_3 - 2 \alpha_4 - \alpha_5 - 2 \alpha_6 \\
      (-1,1,-1,0,0,1) & \mu_{\bf 27} - 2 \alpha_1 - 2 \alpha_2 - 3 \alpha_3 - 2 \alpha_4 - \alpha_5 - \alpha_6 \\
      (-1,1,0,0,0,-1) & \mu_{\bf 27} - 2 \alpha_1 - 2 \alpha_2 - 3 \alpha_3 - 2 \alpha_4 - \alpha_5 - 2 \alpha_6 \\
      (0,-1,0,0,0,1) & \mu_{\bf 27} - 2 \alpha_1 - 3 \alpha_2 - 3 \alpha_3 - 2 \alpha_4 - \alpha_5 - \alpha_6 \\
      (0,-1,1,0,0,-1) & \mu_{\bf 27} - 2 \alpha_1 - 3 \alpha_2 - 3 \alpha_3 - 2 \alpha_4 - \alpha_5 - 2 \alpha_6 \\
      (0,0,-1,1,0,0) & \mu_{\bf 27} - 2 \alpha_1 - 3 \alpha_2 - 4 \alpha_3 - 2 \alpha_4 - \alpha_5 - 2 \alpha_6 \\
      (0,0,0,-1,1,0) & \mu_{\bf 27} - 2 \alpha_1 - 3 \alpha_2 - 4 \alpha_3 - 3 \alpha_4 - \alpha_5 - 2 \alpha_6 \\
      (0,0,0,0,-1,0) & \mu_{\bf 27} - 2 \alpha_1 - 3 \alpha_2 - 4 \alpha_3 - 3 \alpha_4 - 2 \alpha_5 - 2 \alpha_6 \\
  \end{array}
  \label{eq:E6table}
\end{equation}

\end{appendix}

\bibliographystyle{JHEP}

\providecommand{\href}[2]{#2}\begingroup\raggedright\endgroup

\end{document}